\documentclass[a4paper,10pt,twoside]{cpc-hepnp}

\usepackage{multicol}
\usepackage{graphicx}
\usepackage{booktabs}
\usepackage{amssymb,bm,mathrsfs,bbm,amscd}
\usepackage[tbtags]{amsmath}
\usepackage{lastpage}

\begin{document}

%\fancyhead[co]{\footnotesize F. Jone et al: Instruction for typesetting manuscripts}

%\footnotetext[0]{Received 14 March 2009}

\title{Recent results on nucleon resonance electrocouplings from the studies of 
$\pi^{+}\pi^{-}p$ electroproduction with the CLAS detector}

%  \thanks{This work was supported in part by the U.S.Department of Energy and the National Science Foundation,
%  the Skobeltsyn Nuclear Physics Institute and Physics Department at Moscow State University and the Italian
%  Istituto Nazionale di Fisica Nucleare  (55555555) }

\author{%
      V. I. Mokeev$^{1,2}$\email{mokeev@jlab.org}%
\quad V. D. Burkert$^{1}$
\quad L. Elouadrhiri$^{1}$
\\
\quad G. V. Fedotov$^{2}$
\quad E. N. Golovach$^{2}$
\quad B. S.  Ishkhanov$^{2,3}$
  }

 \maketitle

\address{%
1~(Jefferson Lab,  Newport News,  12000 Jefferson Ave, VA 23606,  USA)\\
2~(Skobeltsyn Nuclear Physics Institute at Moscow State University, Moscow 119899, Russia)\\
3~(Physics Department at Moscow State University, Moscow 119899, Russia)\\
%4~(Istituto Nazionale di Fisica Nucleare sez di Genova, Genova via Dodecaneso 33, Italy)\\
}

\begin{abstract}
Recent results on nucleon resonance studies in $\pi^{+}\pi^{-}p$
electro- production off protons with the CLAS detector are presented. 
The analysis of 
CLAS data allowed us to determine all essential
contributing mechanisms, providing a credible separation between resonant 
and non-resonant parts of the cross sections in a wide kinematical area 
of invariant masses of the
final hadronic system $1.3<W<1.8$ GeV and photon virtualities
$0.2<Q^{2}<1.5$ $GeV^2$. Electrocouplings of several excited
proton states with masses less than 1.8 GeV were obtained for the
first time from the analysis of $\pi^{+}\pi^{-}p$ exclusive electroproduction 
channel.
\end{abstract}



\begin{multicols}{2}

\section{Introduction}
Evaluation of electromagnetic $N \rightarrow N^*$ transition
helicity amplitudes ($N^*$ electrocouplings) from the data on
$\pi^{+}\pi^{-}p$ electroproduction represents an important
direction in the studies of $N^*$ structures with CLAS \citep{Bur07}. 
The contributions from  $\pi^{+}n$, $\pi^{0}p$ and
$\pi^{+}\pi^{-}p$ exclusive channels account for almost 90\% of the 
meson electroproduction cross section in the $N^*$ excitation
region. These channels combined are sensitive to a major part of
excited proton states. Moreover, these exclusive channels are strongly coupled by
hadronic interactions of the respective final states. Therefore, nucleon
resonance studies in single (1$\pi$) and charged double pion (2$\pi$) 
exclusive channels are of key importance for the entire $N^*$ physics.

In this proceeding we present recent developments in
phenomenological studies of CLAS data \citep{Ri03,Fe09} on $\pi^{+}\pi^{-}p$ 
electroproduction
off protons. The analysis was carried out in a
wide area of invariant masses of the final hadronic system
$1.3<W<1.8$ GeV and photon virtualities $0.2<Q^{2}<1.5$ $GeV^2$
with the ultimate objective of evaluating  electrocouplings of almost all excited
proton states with masses less than 1.8 GeV from the
fit of all available differential and fully integrated cross
sections combined.
\begin{figure*}[htp]
\begin{center}
\includegraphics[width=16cm]{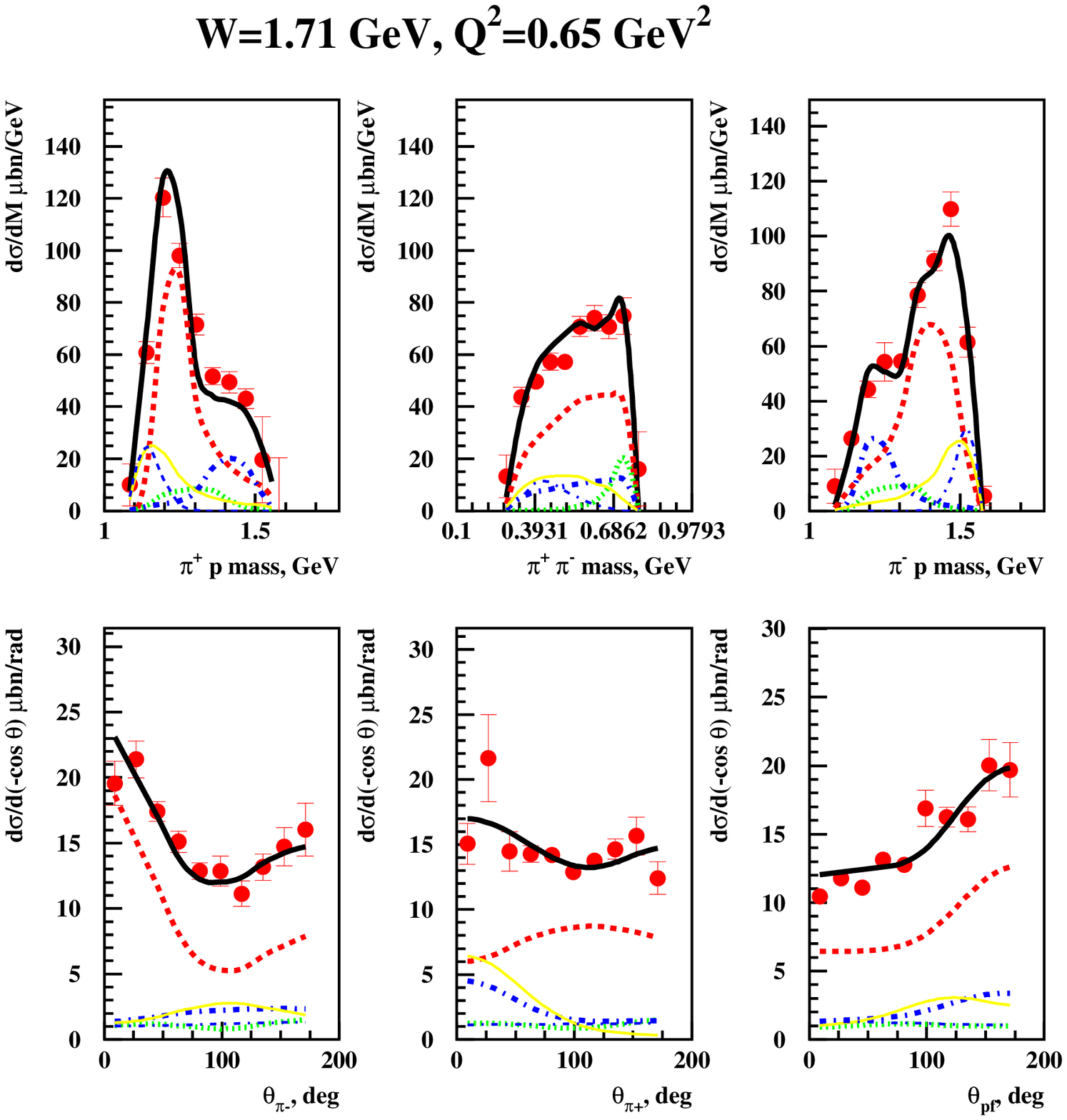}
\caption{Description of the CLAS $\pi^{+}\pi^{-}p$ differential
  cross sections \citep{Ri03} at W = 1.71 GeV and Q$^{2}$ = 0.65~GeV$^{2}$
  within the framework of the updated JM06 model (see Section 2). 
  Full calculations are shown by
  the thick solid lines. Contributions from $\pi^{-} \Delta^{++}$ and
  $\pi^{+} \Delta^{0}$ isobar channels are shown by the thick dashed and
  dash-dotted lines, respectively, and contributions from the $\pi^{+} D_{13}(1520)$, $\rho p$, 
  $\pi^{+} F_{15}(1685)$  isobar channels are shown by the thin solid, dotted and dot-dashed
  lines.}
\label{9sectok}
\end{center}
\end{figure*}

\begin{figure*}[htp]
\begin{center}
\includegraphics[width=16cm]{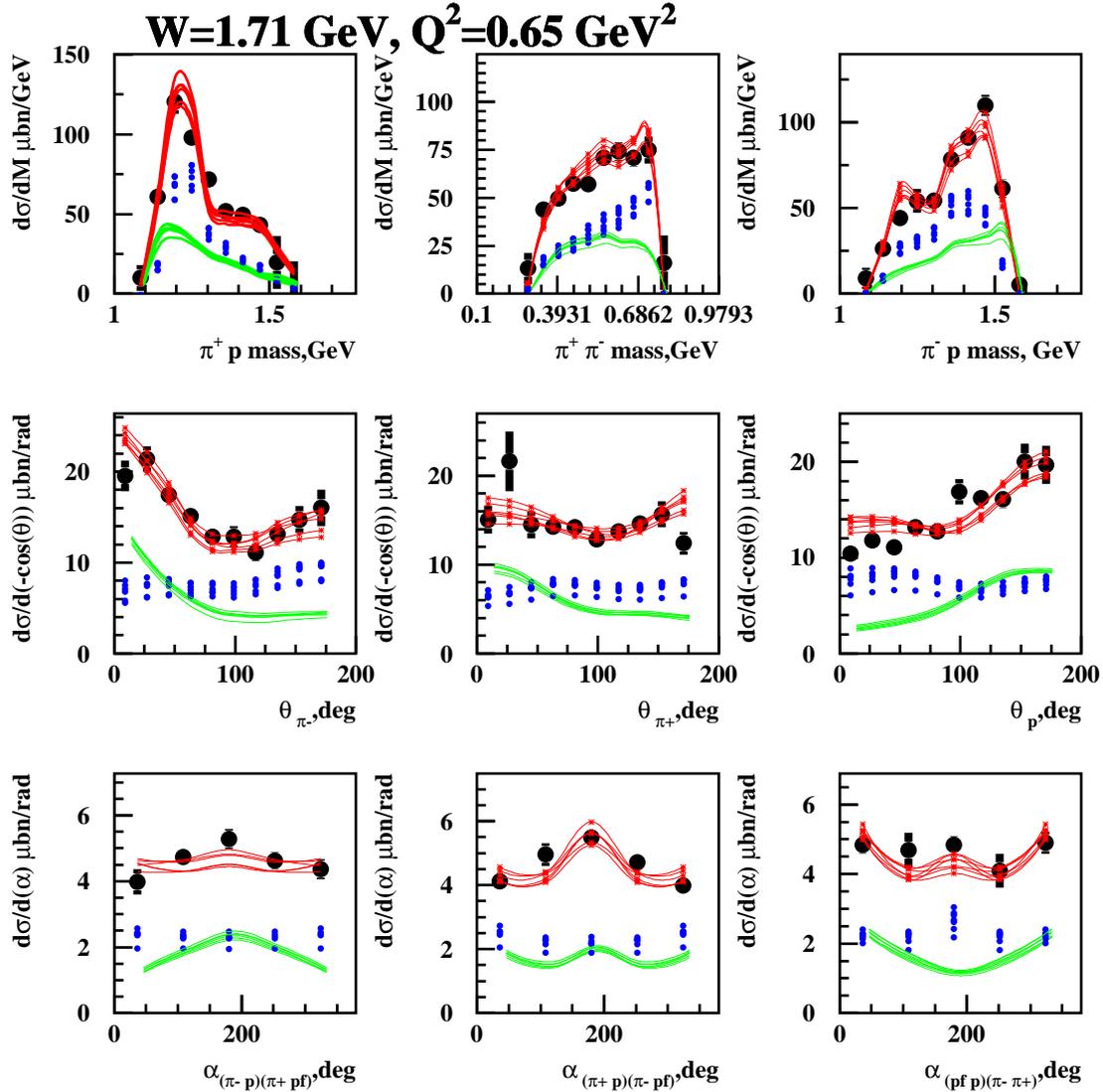}
\caption{Resonant (bars) and non-resonant (lines)
  contributions to the charged double pion differential cross sections
  at $W=1.71$~GeV and $Q^{2}=0.65$~GeV$^{2}$. The full JM
  calculation are shown by solid red lines. 
  The experimental data are from \citep{Ri03}.}
\label{9secnstbck}
\end{center}
\end{figure*}

\begin{figure*}[ht]
\begin{center}
\includegraphics[width=12cm]{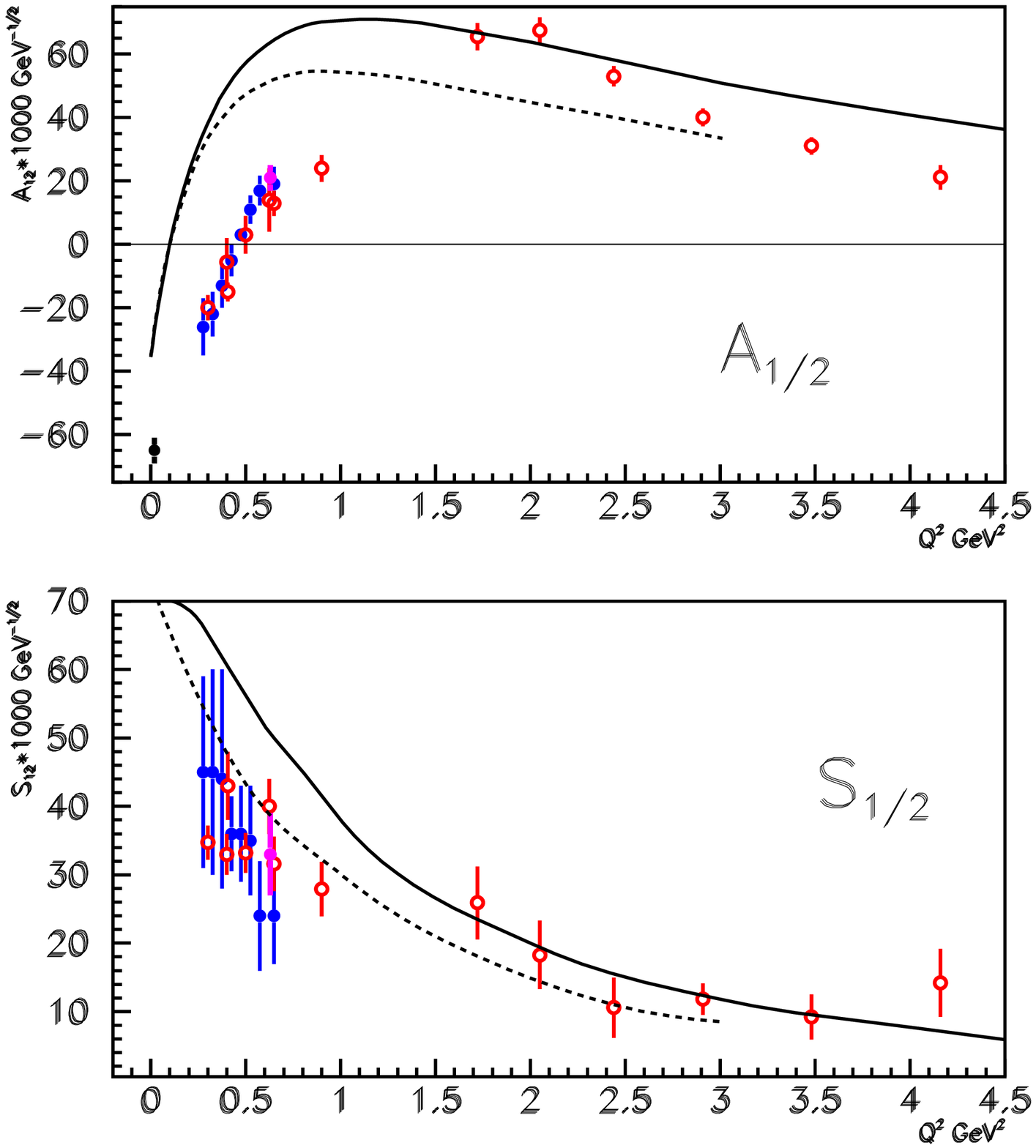}
\caption{Electrocouplings of the $P_{11}(1440)$   
 on the proton in units of $10^{-3}$~GeV$^{-1/2}$. CLAS
  results~\citep{Az05d,Az08d,Bu08c} of the 1$\pi$ production data are 
  represented by the open
  circles. 
  Filled circles are the results of the preliminary 2$\pi$ data~\citep{Mo08m} 
  at low $Q^{2}$ and of the combined
  analysis of the 1$\pi$ and 2$\pi$ channels~\citep{Az05}. Solid and dashed lines are the calculations within the
  framework of the light front
  quark models \citep{Cap95,Az07}.}
\label{p11}
\end{center}
\end{figure*}

\section{Meson-baryon model JM for nucleon resonance studies in 
the $\pi^{+}\pi^{-}p$ exclusive electroproduction
channel}
\label{model} 
The analysis presented in these proceedings incorporates differential and fully
integrated $\pi^{+}\pi^{-}p$  cross sections of  
the recent CLAS data \citep{Fe09}  and previous data \citep{Ri03}. 
These measurements cover a wide  kinematical area of 
1.3$<$W$<$1.8 GeV and 0.25$<$Q$^{2}$$<$1.5
GeV$^2$ providing  for the first time  nine
independent differential cross sections in each bin of $W$ and $Q^{2}$.  
Both experimental data sets \citep{Fe09,Ri03} consist of $\pi^{+}$p, $\pi^{+}$$\pi^{-}$,
$\pi^{-}$p invariant masses;
$\pi^{-}$  $\pi^{+}$ p CM polar angular distributions and of three distributions over angles $\alpha_{i,j}$ 
between two planes, respectively, composed 
by the momenta of
the initial proton and  final hadron (first plane) and two the other
final hadrons (a second plane) with three possible combinations 
amongst these pairs. These data make it possible to establish 
all essential reaction mechanisms 
from the
studies of their
manifestations in observables, as peaks in invariant mass distributions or 
sharp slopes in
angular distributions. The remaining mechanisms without pronounced kinematical
dependencies can be established from the studies of correlations between shapes
of their cross sections in various observables. Employing this strategy the 
phenomenological meson-baryon 
model JM06 \citep{Mo09} 
was developed with primary goal of determining $N^*$
electrocouplings from the combined fit of all nine differential cross sections.

\begin{figure*}[htp]
\begin{center}
\includegraphics[width=5.5cm]{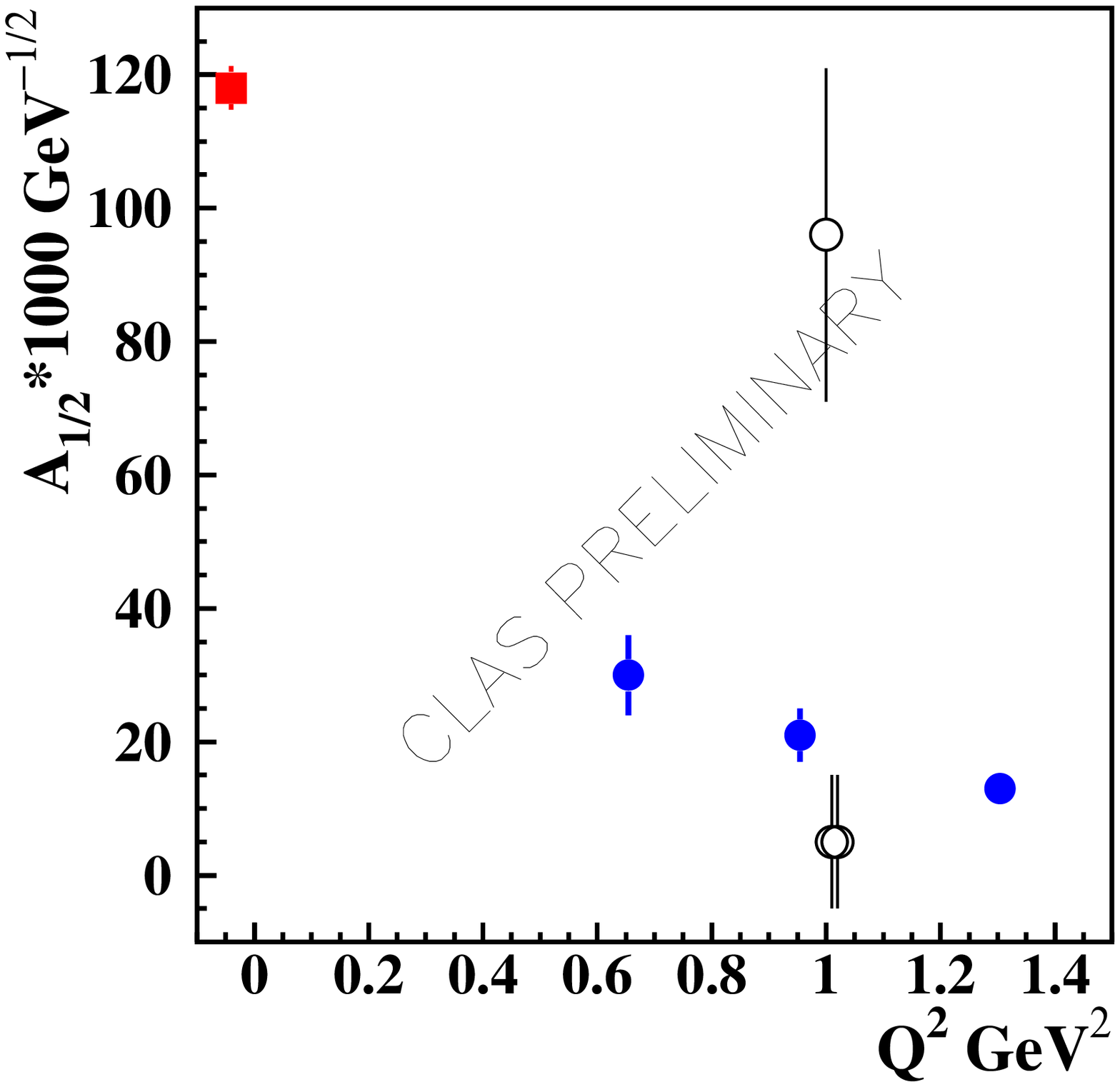}
\includegraphics[width=5.5cm]{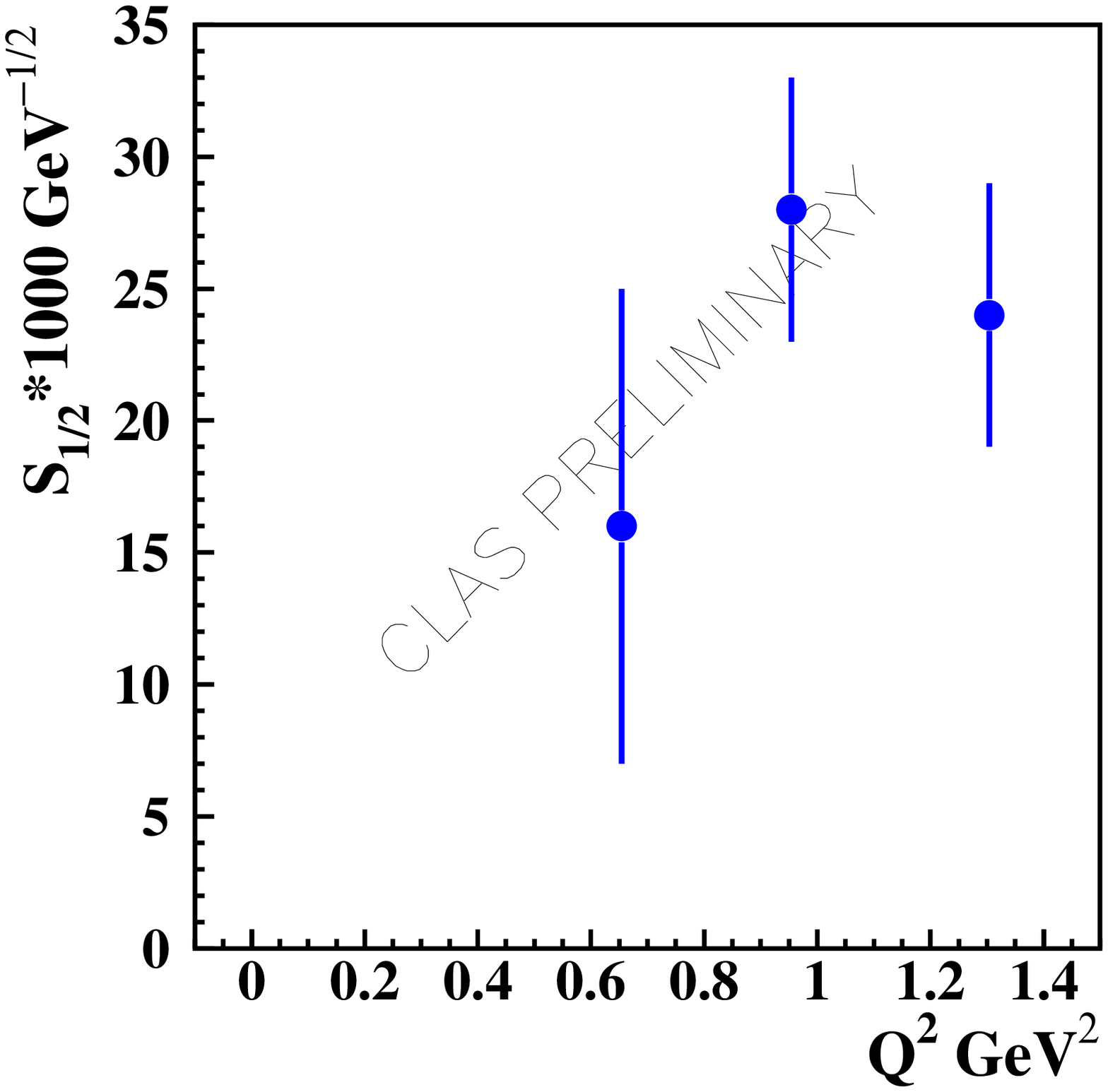}
\includegraphics[width=5.5cm]{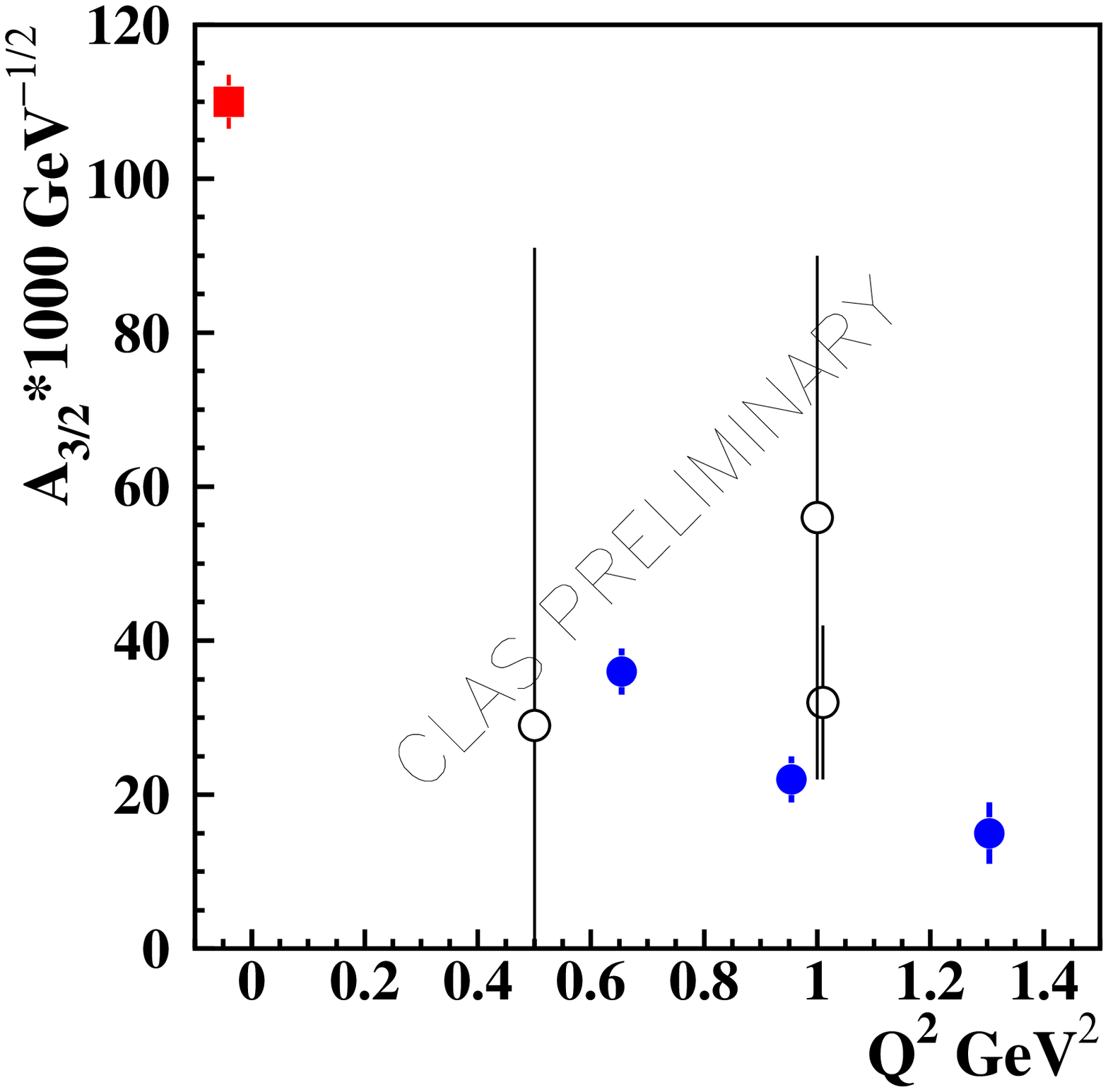}
\caption{Electrocouplings of $D_{33}(1700)$ state from the analysis of CLAS data \citep{Ri03} on $\pi^{+}\pi^{-}p$
electroproduction are shown by filled circles. Open circles are the world data 
from analyses of
single pion electroproduction channels available before CLAS data on 
$\pi^{+}\pi^{-}p$ electroproduction. Photocouplings at $Q^2$=0 GeV$^2$ taken
from \citep{Dug09} are shown by squares.}
\label{d33}
\end{center}
\end{figure*}

This model was initially utilized
for the analysis of data \citep{Fe09} in the kinematic area of 1.3$<$W$<$1.56 GeV and 
0.25$<$Q$^{2}$$<$0.6GeV$^2$. In this kinematic region the JM06 model describes
the $\gamma^* p \rightarrow \pi^+\pi^- p$ 
production amplitude as a superposition of $\pi^-\Delta^{++}$, $\pi^{+} \Delta^{0}$ isobar channels and direct double pion
production mechanisms. The production amplitudes for $\pi \Delta$ intermediate
states consist of resonant contributions 
$\gamma N \rightarrow N^*,\Delta^* \rightarrow \pi\Delta$,
and non-resonant terms. All well established resonances with observed $\pi \Delta$
hadronic decays were incorporated into the JM06 model as well as a 
3/2$^{+}$(1720) candidate state \citep{Ri03}. However, at 1.3$<$W$<$1.56 GeV only the contributions from  
$P_{11}(1440)$ and $D_{13}(1520)$ states were outside of the experimental data uncertainties.
The non-resonant amplitudes were calculated from the
 well established Born terms and 
 presented in  \citep{Ri00,Mo09}. Additional contact terms
were implemented in order to account for possible contributions from other
mechanisms to $\pi \Delta$ production \citep{Mo09}. The analysis of CLAS data 
\citep{Fe09} allowed us to establish the contribution 
from direct charged double pion
 (2$\pi$) production mechanisms, when the $\pi^{+} \pi^{-} p$ final state 
 is created 
without the formation
of unstable hadrons in the intermediate states. The amplitudes for these
processes, have been determined for the first time, are described in \citep{Mo09}.

Recently the JM06 model was updated based on the analysis of 
the CLAS data \citep{Ri03} collected in a wider kinematic area of 1.4$<$W$<$1.8 GeV 
and 0.5$<$$Q^{2}$$<$1.5 $GeV^2$.  
The previous analysis of this data  within the framework of the 
JM05 model version \citep{Mo06-1,Mo06} was limited to four differential  
cross sections: $\pi^{+}$p, $\pi^{+}$$\pi^{-}$,
$\pi^{-}$p invariant masses and
$\pi^{-}$ angular distributions. The current analysis incorporates 
all nine differential cross
sections mentioned above. The JM06 model production mechanisms, determined from the data 
\citep{Fe09} at W$<$1.56 GeV, were extended by implementing quasi-two-body channels: $\rho^{0} p$, 
$\pi^{+} D^0_{13}(1520)$, $\pi^{+} F^0_{15}(1520)$, $\pi^{-} P^{++}_{33}(1640)$.
    
The $\rho^{0} p$ isobar  channel becomes visible in the data at $W > 1.65$~GeV
with significant resonant contributions for  $W < 2.0$~GeV. The production 
amplitudes for $\rho^{0} p$ intermediate
states consist of the resonant contributions 
$\gamma N \rightarrow N^*, \: \; \Delta^* \rightarrow \rho^{0} p$,
and non-resonant terms. Here the
non-resonant amplitudes are estimated by a diffractive ansatz, that
has been modified in order to reproduce experimental data in the near
and sub-threshold regions \citep{Shv07}.

 The contributions from the $\pi^{+} D_{13}^{0}(1520)$ and $\pi^{+}
F_{15}^{0}(1685)$  isobar channels are
seen in the $\pi^{-} p$  and $\pi^{+} p$ mass distributions at $W >
1.65$~GeV (Fig.~\ref{9sectok}). The $\pi^{+} D_{13}^{0}(1520)$ amplitudes are derived from
the Born terms of the $\pi \Delta$ isobar channels by implementing an
additional $\gamma_{5}$-matrix that accounts for the opposite parity
of the $D_{13}(1520)$  with respect to the $\Delta$. The amplitudes of 
$\pi^{+} F_{15}^{0}(1685)$  isobar
channel are parametrized by Lorentz invariant contractions of the
initial and final particle spin-tensors and by effective propagators
for the intermediate state particles. The magnitudes of these
amplitudes are fitted to the data \citep{Mo06}.

Within the framework of the updated JM06 approach, we achieved a good description
of the CLAS $\pi^{+}\pi^{-}p$ data over the entire kinematic range: 1.31$<$W$<$1.8 GeV and
0.25$<$$Q^2$$<$1.5 GeV$^2$. As a typical example, the model description of the nine
differential cross sections at $W = 1.71$~GeV and $Q^{2} =
0.65$~GeV$^{2}$ are presented in Fig.~\ref{9sectok} together with
the contributing mechanisms. Direct 2$\pi$ electroproduction mechanisms, that 
account for up to 30\% of the cross sections
at W$<$ 1.56 GeV \citep{Mo09}, decreases with W and become negligible at 
W$>$ 1.65 GeV. The different mechanisms result in
qualitatively different shapes of their respective contributions to 
various observables. The successful simultaneous description of all nine
differential cross sections enables us to identify the 
essential contributing processes and to access their dynamics at the
phenomenological level.

\begin{figure*}[htp]
\begin{center}
\includegraphics[width=5.5cm]{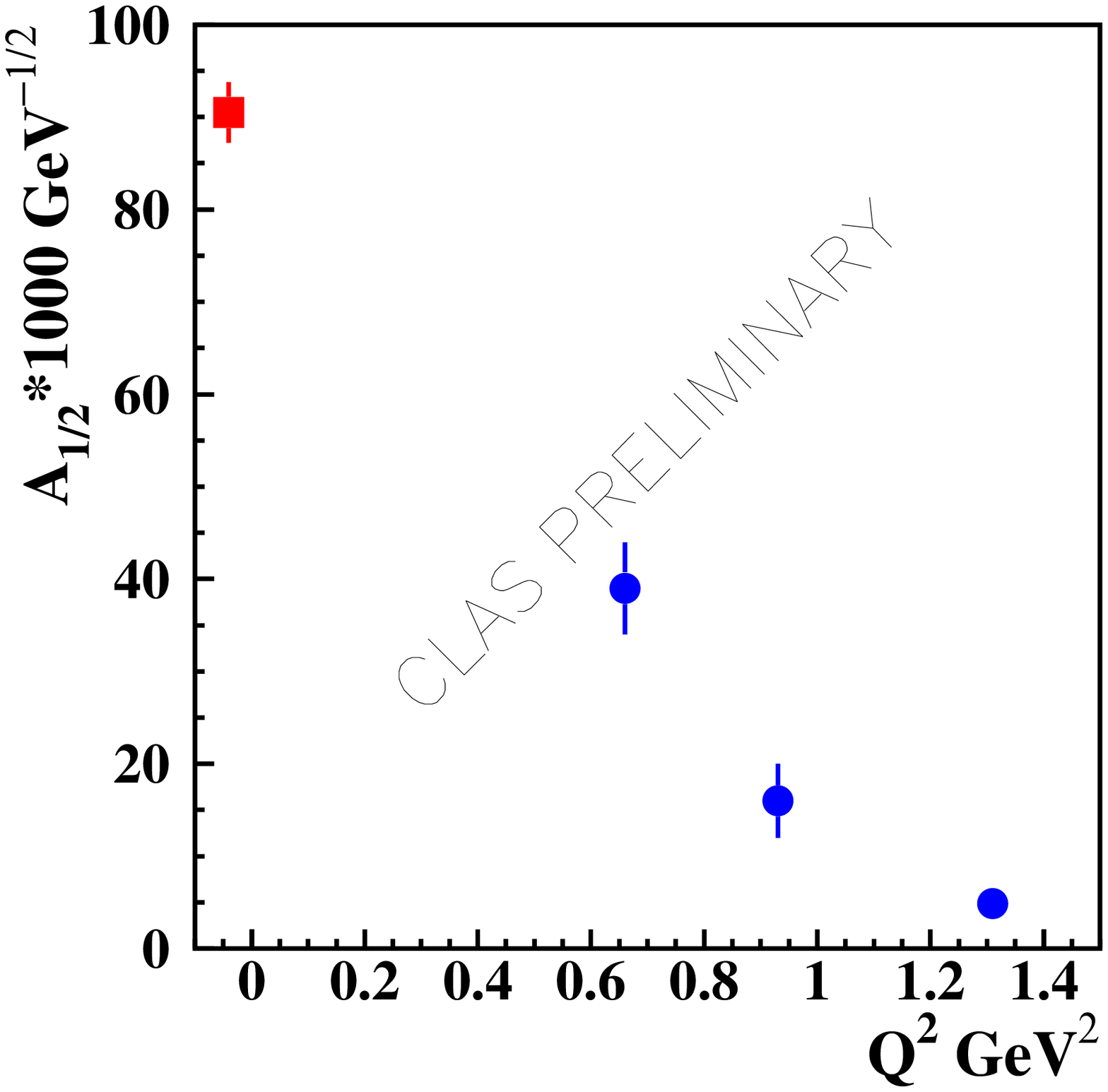}
\includegraphics[width=5.5cm]{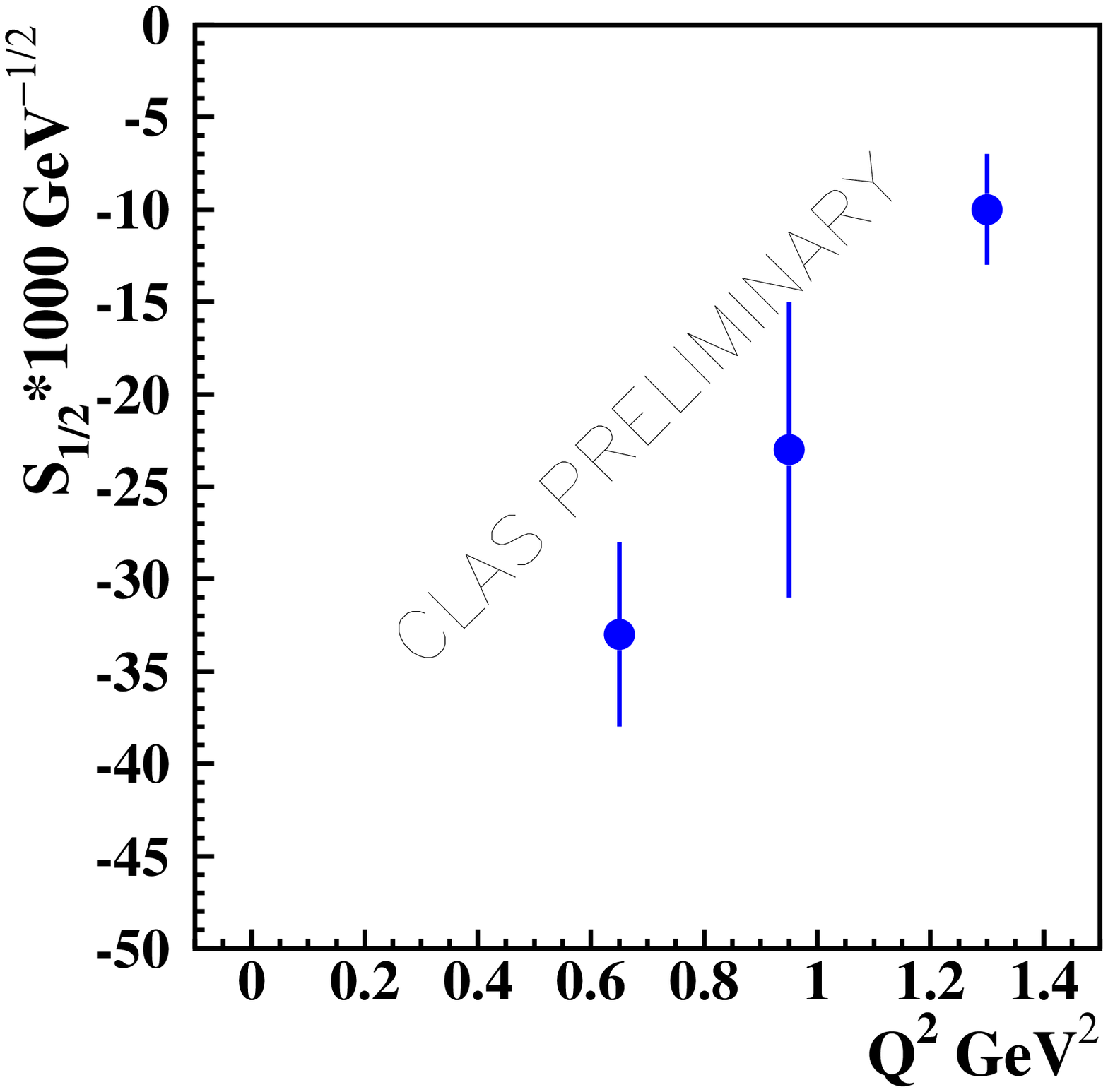}
\includegraphics[width=5.5cm]{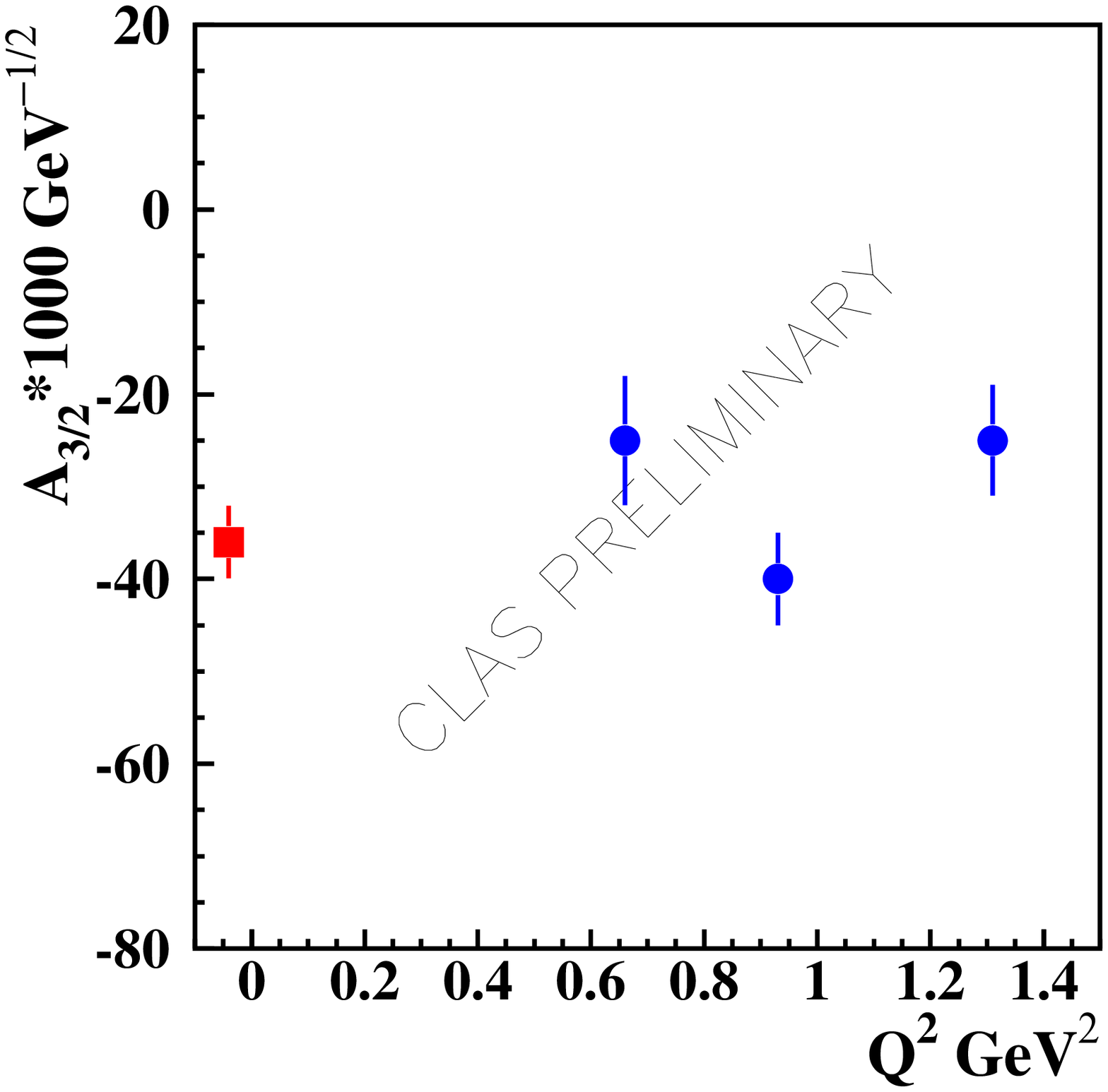}
\caption{Electrocouplings of $P_{13}(1720)$ state from analysis of the CLAS data \citep{Ri03} on $\pi^{+}\pi^{-}p$
electroproduction are shown by filled circles. 
Photocouplings at $Q^2$=0 GeV$^2$ taken from \citep{Dug09} are shown by squares.}
\label{p13}
\end{center}
\end{figure*}

\section{Resonance electrocouplings from the $\pi^{+}\pi^{-}p$ 
electroproduction}
 The separation of resonant and non-resonant contributions based on the
JM model parameters are shown in Fig.~\ref{9secnstbck}. Resonant and 
non-resonant parts have
qualitatively different shapes in all observables. This allows us to
isolate the resonant contributions unambiguously. Uncertainties for resonant and non-resonant parts of cross
sections shown in Fig.~\ref{9secnstbck} are comparable with uncertainties of experimental data. 
Good separation between resonant and non-resonant cross sections, achieved 
within the framework of the JM model, 
allowed us to extract the $N^{*}$
electrocouplings for several excited proton states with masses less than 1.8 GeV.

The CLAS data have enabled us for the first time to determine 
the $P_{11}(1440)$,  electrocouplings over a wide range of photon virtualities 
by analyzing the
two major exclusive channels: 1$\pi$ and 2$\pi$ electroproduction
\citep{Fe09,Ri03,Az05d,Az08d,Bu08c}. As an example, the
electrocouplings of the $P_{11}(1440)$ 
state are shown in
Fig.~\ref{p11}.

The agreement of the results 
obtained from the analyses of 1$\pi$ and 2$\pi$ channels is highly 
significant since these meson electroproduction
channels have completely different non-resonant amplitudes. The
successful description of the large body of data on
1$\pi$ and 2$\pi$ electroproduction with almost the same values 
for the $P_{11}(1440)$  electrocouplings,
shows the capability of the analysis methods developed for description of 1$\pi$
and 2$\pi$ electroproduction to provide a
reliable evaluation of the resonance parameters.

Light cone quark models \citep{Cap95,Az07} provided reasonable data description at $Q^2$ $>$ 2.0
GeV$^2$ and, therefore, strongly suggest that the quark core of 
$P_{11}(1440)$ state represents predominantly the first radial excitation of 
the three quark ground state nucleon. For the first time we observed the sign
change in the Q$^2$ evolution of
$A_{1/2}$ electrocoupling for $P_{11}(1440)$ state. This distinctive feature 
was predicted by light cone quark models,
showing the essential role of light front effects in the structure of $P_{11}(1440)$. 
However, the light cone model expectations \citep{Cap95,Az07} and the data on $A_{1/2}$ electrocouplings of $P_{11}(1440)$ become considerably
different at $Q^2$$<$1.0 GeV$^2$. This discrepancy offers an indication that at
distances of the order of the typical hadron size, not only quark core, but also others degrees of freedom as
meson-baryon dressing may have substantial contribution 
to the resonance structure. Extensive studies of
meson-baryon dressing contributions to the structure of low lying $N^*$ are now
in progress \citep{Lee07}.

The CLAS data on $\pi^{+}\pi^{-}p$
electroproduction \citep{Ri03} allowed us for the first time to determine 
with a good accuracy electrocouplings of
high lying resonances, that mostly decay to the $N\pi\pi$ final states.
These data analysis 
was carried
out within the framework of the updated JM model, outlined above. 
N$^*$ hadronic couplings were taken from analyses of experiments with hadronic
probes, presented in PDG. As it was shown in
\citep{Ri03}, with  this choice of resonance hadronic couplings we need 
to account for the contributions from $3/2^{+}(1720)$  candidate state.
Electrocouplings of $D_{33}(1700)$ and $P_{13}(1720)$ states obtained from the
analysis of  the $\pi^{+}\pi^{-}p$ electroproduction data \citep{Ri03} 
within the framework of the JM
model are shown in Fig.~\ref{d33} and Fig.~\ref{p13}. For comparison the world
data on electrocouplings of $D_{33}(1700)$ state,
available before the CLAS $\pi^{+}\pi^{-}p$ electroproduction data, are
shown by open circles. They were determined from the analysis of single pion
electroproduction channels. The branching fraction to the
$N\pi$ final states for the $D_{33}(1700)$ decays  is less than 20\%. 
Therefore, single pion electroproduction
channels have not enough sensitivity to the electrocouplings of this state. 
This is the reason for huge uncertainties of the previous world data shown in 
Fig.~\ref{d33}.
The studies of $\pi^{+}\pi^{-}p$ electroproduction provided first information
on the electrocouplings of $D_{33}(1700)$ and $P_{13}(1720)$ states with reasonable
accuracy. We observed a rapid fall-off of the $A_{1/2}$ electrocoupling of 
$P_{13}(1720)$ state with the photon
virtuality Q$^2$. At Q$^2$ above 0.9 GeV$^2$ the absolute values of the helicity 
non-conserving $A_{3/2}$
electrocouplings become larger than the helicity conserving $A_{1/2}$ amplitude.
 Search for the helicity conserving regime,
expected at asymptotically high photon virtualities, represents an interesting 
open question for the future 
studies of this state at high photon virtualities.

\section{Conclusions}
Analysis of the CLAS data on the charged double pion electroproduction off 
protons \citep{Fe09,Ri03} within the framework of the JM model allowed us to establish all essential
mechanisms contributing to this exclusive channel at 1.31$<$W$<$1.8 GeV and
0.25$<$$Q^2$$<$1.5 GeV$^2$. A good description of nine differential cross sections in each bin of W and Q$^2$ was
achieved, allowing us to isolate the resonant contributions to cross sections, needed for evaluation of $N^*$
electrocouplings. For the first time electrocouplings of several excited proton states with masses less than 1.8
GeV were determined from the $\pi^{+}\pi^{-}p$ electroproduction channel. 
Electrocouplings of $P_{11}(1440)$ state
obtained from the analysis of two major single and charged double pion 
electroproduction channels with completely
different non-resonant mechanisms are in a good agreement, showing 
the capability of the JM model to provide a reliable evaluation of 
the resonance parameters. The studies of $\pi^{+}\pi^{-}p$ electroproduction provided 
first  data of reasonable accuracy on electrocouplings of the high lying resonances $D_{33}(1700)$ and $P_{13}(1720)$, that mostly decay with two
pion emission.
\vspace{0.5cm}
\\
\acknowledgments{Acknowledgments.\\
This work was supported in part by the U.S.Department of Energy and the National Science Foundation,
  the Skobeltsyn Nuclear Physics Institute and Physics Department 
  at Moscow State University. Jefferson Science Associates (JSA) operates the
  Thomas Jefferson National Accelerator Facility for the U.S.Department of
  Energy under contract DE-AC05-060R23177.}

\end{multicols}

\begin{multicols}{2}

\end{multicols}

\clearpage

\end{document}